\documentclass[twocolumn]{aastex631}

\usepackage[latin1]{inputenc}

\usepackage{amsmath}
\usepackage{bm}
\usepackage{tensor}
\usepackage{graphicx}
\usepackage{hyperref}
\usepackage[noabbrev,capitalise]{cleveref}

\newcommand{\ii}{\mathrm{i}}
\newcommand{\fg}{\omega_{\mathrm{g}}}
\newcommand{\fe}{\omega_{\mathrm{e}}}
\newcommand{\fho}{\omega_{\mathrm{h}}}
\newcommand{\fz}{\omega_{0}}
\newcommand{\ip}{e^{\ii P}}

\newcommand{\ir}{e^{\ii R}}
\newcommand{\id}{e^{\ii\Delta}}
\newcommand{\ipq}{e^{\ii\left(P+Q\right)}}
\newcommand{\ipqd}{e^{\ii\left(P+Q+\Delta\right)}}
\newcommand{\intv}{\int_{\mathcal{V}}d\mathcal{V}}

\newcommand{\eps}[2]{\tensor[^{\epsilon}]{#1}{#2}}

\begin{document}

\title{Extending the Observational Frequency Range for Gravitational Waves in a Pulsar Timing Array}

\author[0000-0002-2692-7520]{Chan Park}
\affiliation{National Institute for Mathematical Sciences, Daejeon 34047, Republic of Korea}

\begin{abstract}
We provide an observation method for gravitational waves using a pulsar timing array to extend the observational frequency range up to the rotational frequency of pulsars. For this purpose, we perform an analysis of a perturbed electromagnetic wave in perturbed spacetime from the field perspective. We apply the analysis to the received electromagnetic waves in a radio telescope, which partially composes the periodic electromagnetic pulse emitted by a pulsar. For simple observation, two frequency windows are considered. For each window, we propose gauge-invariant quantities and discuss their observations.
\end{abstract}

\section{Introduction}
A pulsar is a stably rotating compact star that periodically emits an electromagnetic pulse (EMP) due to its magnetic axis being tilted to the rotation axis. Because the rotation of a pulsar is extremely stable, the period of the EMP generally remains roughly constant. However, when an EMP passes through a perturbed spacetime, its period will be changed, and the perturbation will contain information about gravitational waves (GWs).

In practice, GW observation by a single pulsar measurement is not possible because various noises are larger than the GW signal. Instead, the correlation between measurements from different pulsars can amplify a stochastic gravitational wave (SGW) signal by increasing the measurement time if the noises of the measurements are independent of each other. In this way, observations are being performed to detect the SGW for the first time by pulsar timing arrays (PTAs) (\cite{shannon_gravitational_2015,babak_european_2016,arzoumanian_nanograv_2020}). The method of extracting an SGW signal from the correlation was proposed by \cite{hellings_upper_1983}.

Let us look at the analysis for the effect of GW on the EMP period. When we consider an electromagnetic wave (EMW) in a perturbed spacetime with GWs such that the frequency of the EMW is much larger than those of the GWs, we can introduce geometrical optics. By an approximation of geometrical optics, the wavevector of the EMW is null and follows null geodesic (\cite{misner2017gravitation}). From the solution of the perturbed null geodesic equation with appropriate boundary conditions, we obtain information of the perturbed EMW, e.g., redshift and angular deflection as in \cite{book_astrometric_2011}.

On the other hand, the time interval between emission and arrival time of the EMP is obtained by the perturbation of the null condition as in \cite{Maggiore2018}. Additionally, assuming that GW frequency is much smaller than the EMP frequency, we obtain the well-known redshift formula given by \cite{estabrook_response_1975} and \cite{detweiler_pulsar_1979}. Note that we distinguish between the frequency of EMP and EMW. The EMP frequency is defined from the periodic intensity of the EMWs, which is quadratic to the electromagnetic field. Therefore, a periodic EMP can be composed of EMWs with relatively high frequencies. For example, a radio telescope with a GHz frequency band of EMW measures a periodic EMP in KHz.

In this paper, we determine an analysis of perturbed EMWs that applies to the observation of GWs with frequencies comparable to the EMP frequency of a pulsar. To do this, we first obtain a general solution of the vacuum Maxwell equation in the presence of GWs without using the geometric optical approximation. The general solution consists of a particular solution and a homogeneous solution. The particular solution is determined by GWs, and the homogeneous solution is determined to satisfy the appropriate boundary condition for the electromagnetic field. In \cite{montanari_propagation_1998}, they performed an analysis of EMWs with GWs by solving the Maxwell equation introducing the Fermi normal coordinate. Instead, we will give a covariant analysis without introducing a coordinate system.

For measurement of EMWs, we introduce spatial scalars of the electromagnetic field independent of the spatial frame in a given observer, e.g., $E^{2}$, $B^{2}$, and $E\cdot B$. When the measurement is given in components that depend on a spatial frame, the analysis of its perturbation is challenging because we need the frame perturbation that reflects the structure of the device. In contrast, spatial scalars are given only by the electric field, the magnetic field, and the spatial metric.

Finally, we restrict frequencies of GWs to much less than that of EMWs to introduce the geometrical optics approximation. On the other hand, the frequency of a GW can be comparable to that of an EMP. With these GWs, we obtain the perturbed spatial scalars of a periodic EMP composed by high frequency EMWs. As an application, two frequency windows for spatial scalars are considered. For each window, we propose gauge-invariant quantities and discuss their observations.

In this paper, we introduce the geometrized unit $\left(c=1,G=1\right)$ for the spacetime, and the Gaussian unit $\left(\epsilon_{0}=1/4\pi,\mu_{0}=4\pi\right)$ for the electromagnetism. Indices represented by italic lowercase Latin letters starting from $a$ are abstract indices, as in \cite{wald1984general}.

\section{Ansatz}
\label{sec:ansatz}

We consider a one-parameter family of perturbed spacetime $\left(M_{\epsilon},g\left(\epsilon\right)\right)$ where $\epsilon$ is a perturbation parameter and $\left(M_{0},g\right)$ is the Minkowski background spacetime. We need a diffeomorphism $\phi_{\epsilon}$ to identify points between $M_{0}$ and $M_{\epsilon}$. A perturbed quantity is defined by the pull-back of a quantity from $M_{\epsilon}$ to $M_{0}$ denoted with the left superscript of $\epsilon$, e.g., $\eps{Q}{}\equiv\phi_{\epsilon}^{*}Q\left(\epsilon\right)$ for a geometrical quantity $Q\left(\epsilon\right)$. This style of perturbation in a coordinate invariant manner was introduced in \cite{stewart1974} and \cite{stewart1990}.

Perturbed metric $\eps{g}{}$ is expanded as Taylor series by
\begin{align}
    \eps{g}{_{ab}}=g_{ab}+\epsilon h_{ab}+O\left(\epsilon^{2}\right),
\end{align}
where $g$ is the Minkowski background metric, and $h$ is the first-order metric perturbation. We impose gauge conditions on $h$ as
\begin{align}
    \nabla^{b}h_{ab}&=0,
    \\\tensor{h}{^{a}_{a}}&=0,
\end{align}
where $\nabla$ is the Levi-Civita connection associated with $g$. Then, the perturbation of the vacuum Einstein equation becomes the wave equation
\begin{align}
    \nabla^{c}\nabla_{c}h_{ab}=0.
\end{align}

Let us consider geodesic observers with four-velocity vector field $\eps{n}{}$ over perturbed spacetime given by
\begin{align}
    \eps{n}{^{a}}=n^{a}+\epsilon\delta{n}^{a}+O\left(\epsilon^{2}\right),
\end{align}
where $n^{a}\equiv-g^{ab}\nabla_{b}t$ is the normal vector of globally inertial time coordinate $t$ in Minkowski spacetime, and $\delta n$ is the first-order perturbation. Then, the perturbed geodesic equation becomes
\begin{align}
    \delta n^{b}\nabla_{b}n^{a}+n^{b}\nabla_{b}\delta n^{a}+\tensor{C}{^{a}_{bc}}n^{b}n^{c}=0,\label{eq:geodesic_perturbation}
\end{align}
where $\tensor{C}{^{a}_{bc}}\equiv\frac{1}{2}\left(\nabla_{b}\tensor{h}{^{a}_{c}}+\nabla_{c}\tensor{h}{^{a}_{b}}-\nabla^{a}h_{bc}\right)$. Additionally imposing a gauge condition on $h$ as
\begin{align}
    h_{ab}n^{b}=0,
\end{align}
we obtain $n^{b}\nabla_{b}\delta n^{a}=0$ from \cref{eq:geodesic_perturbation}. Assuming observers with $\delta n^{a}=0$ before arrival of GW, their four-velocity perturbations remain as $\delta n^{a}=0$. We only consider these observers. Note that the observers of our choice are identical to the observers with fixed TT coordinate (\cite{misner2017gravitation}).

Metric perturbation $h$ is superposed by monochromatic plane GW over all directions and all frequencies. It is given by
\begin{align}
    h_{ab}\left(t,\bm{x}\right)=\int d^{2}\kappa\int_{-\infty}^{\infty}\frac{d\fg}{2\pi}\tilde{h}_{ab}\left(\fg,\kappa\right)e^{\ii P\left(t,\bm{x};\fg,\kappa\right)},\label{eq:h_Fourier}
\end{align}
where $\kappa$ is the spatial unit vector, $d^{2}\kappa$ is the solid angle element of three-dimensional unit sphere, $\fg$ is the frequency, $\tilde{h}$ is the amplitude of GW with $\left(\fg,\kappa\right)$, and $P\left(t,\bm{x};\fg,\kappa\right)\equiv\fg\left(-t+\kappa\cdot\bm{x}\right)$ is the phase. Note that \cref{eq:h_Fourier} does not cover all solutions of $h$. For example, gravitational memory effect (\cite{zeldovich_radiation_1974,smarr_gravitational_1977}) can not be expressed in the form of \cref{eq:h_Fourier}. Let $\tilde{h}$ be zero when $\fg=0$. This corresponds to a gauge choice of $h$. Wavevector is defined by
\begin{align}
    k^{a}\left(\fg,\kappa\right)\equiv g^{ab}\nabla_{b}P,
\end{align}
having 3+1 decomposition (\cite{Gourgoulhon2012}) as
\begin{align}
    k^{a}=\fg\left(n^{a}+\kappa^{a}\right).
\end{align}

We assume that the electromagnetic potential $\eps{A}{}$ has first-order strength in $\epsilon$, i.e.,
\begin{align}
    \eps{A}{_{a}}=\epsilon A_{a}+\epsilon^{2}X_{a}+O\left(\epsilon^{3}\right),
\end{align}
where $A$ is leading order value, and $X$ is its next order perturbation. In the field strength,
\begin{align}
    \eps{F}{_{ab}}&=2\nabla_{[a}\eps{A}{_{b]}}
    \nonumber\\&=\epsilon F_{ab}+\epsilon^{2}Y_{ab}+O\left(\epsilon^{3}\right),
\end{align}
where $F_{ab}=2\nabla_{[a}A_{b]}$ and $Y_{ab}=2\nabla_{[a}X_{b]}$. Additionally, we assume no electromagnetic source. Then, its stress energy
\begin{align}
    \eps{T}{_{ab}}&=\frac{1}{4\pi}\left(\eps{F}{_{a}^{c}}\eps{F}{_{bc}}-\frac{1}{4}g_{ab}\eps{F}{_{cd}}\eps{F}{^{cd}}\right)
    \nonumber\\&=\epsilon^{2}T_{ab}+O\left(\epsilon^{3}\right),
\end{align}
having $O\left(\epsilon^{2}\right)$ does not influence metric perturbation $h$ by the perturbed Einstein equation. We summarize perturbation orders of major quantities in \cref{tab:perturbation}.

\begin{deluxetable}{ccccc}
    \tablecaption{Perturbation orders for major quantities. \label{tab:perturbation}}
    \tablehead{\colhead{Perturbation} & \colhead{Spacetime} & \colhead{Field} & \colhead{Stress} & \colhead{Spatial} \\
    \colhead{Order} & \colhead{Metric} & \colhead{Strength} & \colhead{Energy} & \colhead{Scalars}}
    \startdata
    Zeroth & $g$      & $0$      & $0$      & $0$           \\
    First & $h$      & $A, F$   & $0$      & $0$           \\
    Second & $\cdots$ & $X, Y$   & $T$      & $I_{i}$       \\
    Third & $\cdots$ & $\cdots$ & $\cdots$ & $\delta I_{i}$\\
    \enddata
    \tablecomments{We impose a condition in which stress energy has second-order strength. Then, by the Einstein equation, the Minkowski spacetime metric $g$ in the zeroth-order is compatible to the zeroth-order vacuum, and the metric perturbation $h$ in the first-order is not influenced by the stress energy.}
\end{deluxetable}

We impose gauge conditions on $A$ as
\begin{align}
    \nabla^{a}A_{a}&=0,
    \\A_{a}n^{a}&=0.
\end{align}
Then, the Maxwell equation in leading order becomes the wave equation
\begin{align}
    \nabla^{b}\nabla_{b}A_{a}=0.
\end{align}
We consider a plane EMW propagating to spatial unit vector $\lambda$ as
\begin{align}
    A_{a}\left(t,\bm{x}\right)=\int_{-\infty}^{\infty}\frac{d\fe}{2\pi}\tilde{A}_{a}\left(\fe\right)e^{\ii Q\left(t,\bm{x};\fe,\lambda\right)},\label{eq:4_potential}
\end{align}
where $\fe$ is the frequency, $\tilde{A}$ is the amplitude of EMW with $\left(\fe,\lambda\right)$, and $Q\left(t,\bm{x};\fe,\lambda\right)\equiv\fe\left(-t+\lambda\cdot\bm{x}\right)$ is the phase. Let $\tilde{A}$ be zero when $\fe=0$. This corresponds to a gauge choice of $A$. Wavevector is defined by
\begin{align}
    l^{a}\left(\fe,\lambda\right)\equiv g^{ab}\nabla_{b}Q,
\end{align}
having 3+1 decomposition as
\begin{align}
    l^{a}=\fe\left(n^{a}+\lambda^{a}\right).
\end{align}

For measurement of an EMW, we introduce basic spatial scalars defined by
\begin{align}
    \eps{I}{_{1}}&\equiv\eps{\gamma}{^{ab}}\eps{E}{_{a}}\eps{E}{_{b}}, \\\eps{I}{_{2}}&\equiv\eps{\gamma}{^{ab}}\eps{B}{_{a}}\eps{B}{_{b}}, \\\eps{I}{_{3}}&\equiv\eps{\gamma}{^{ab}}\eps{E}{_{a}}\eps{B}{_{b}},
\end{align}
where $\eps{E}{_{a}}=\eps{F}{_{ab}}\eps{n}{^{b}}$ is the electric field, $\eps{B}{_{a}}=\frac{1}{2}\eps{\varepsilon}{^{bc}_{a}}\eps{F}{_{bc}}$ is the magnetic field, $\eps{\gamma}{}$ is the spatial metric, and $\eps{\varepsilon}{}$ is the spatial Levi-Civita tensor. All spatial scalars that are composed only by electromagnetic quantities $\left(\eps{E}{},\eps{B}{}\right)$ and spacetime quantities $\left(\eps{\gamma}{},\eps{\varepsilon}{}\right)$ can be expressed by a combination of $\eps{I}{_{1}}$, $\eps{I}{_{2}}$, and $\eps{I}{_{3}}$. For example, the magnitude of the Poynting vector is given by $\frac{1}{4\pi}\sqrt{\eps{I}{_{1}}\eps{I}{_{2}}-\eps{I}{_{3}^{2}}}$. The basic spatial scalars are expanded in $\epsilon$ as
\begin{align}
    \eps{I}{_{i}}=\epsilon^{2}I_{i}+\epsilon^{3}\delta I_{i}+O\left(\epsilon^{4}\right),
\end{align}
for $i\in\left\{1,2,3\right\}$.

\section{Perturbation of EMW} 

The perturbation of electromagnetic potential $X$ is determined by the perturbed Maxwell equation, i.e.,
\begin{align}
    \nabla^{b}\nabla_{b}X_{a}&=2\tensor{C}{^{c}_{a}^{b}}\nabla_{b}A_{c}+h^{bc}\nabla_{b}\nabla_{c}A_{a}
    \nonumber\\&=\intv\left(2\ii\tensor{\tilde{C}}{^{c}_{ab}}l^{b}\tilde{A}_{c}-\tilde{h}_{bc}l^{b}l^{c}\tilde{A}_{a}\right)\ipq,\label{eq:eq_of_X}
\end{align}
where $\tensor{\tilde{C}}{^{a}_{bc}}\equiv\frac{1}{2}\ii\left(k_{b}\tensor{\tilde{h}}{^{a}_{c}}+k_{c}\tensor{\tilde{h}}{^{a}_{b}}-k^{a}\tilde{h}_{bc}\right)$, and the integration over domain $\mathcal{V}$ is defined by
\begin{align}
    \intv\equiv\int_{\kappa\neq\lambda} d^{2}\kappa\int_{\substack{-\infty\\\fg\neq0}}^{\infty}\frac{d\fg}{2\pi}\int_{\substack{-\infty\\\fe\neq0}}^{\infty}\frac{d\fe}{2\pi},
\end{align}
excluding the case of $\kappa^{a}=\lambda^{a}$ from $\mathcal{V}$, in which the right-hand side of \cref{eq:eq_of_X} vanishes. The solution $X$ consists of homogeneous solution $X^{\mathrm{p}}$ and particular solution $X^{\mathrm{h}}$. As shown in \cite{park_observation_2021}, the particular solution is obtained by
\begin{align}
    X^{\mathrm{p}}_{a}=\intv\tilde{X}^{\mathrm{p}}_{a}\ipq,
\end{align}
where $\tilde{X}^{\mathrm{p}}$ is
\begin{align}
    \tilde{X}^{\mathrm{p}}_{a}=\frac{1}{k^{d}l_{d}}\left(-2\ii\tensor{\tilde{C}}{^{c}_{ab}}l^{b}\tilde{A}_{c}+\tilde{h}_{bc}l^{b}l^{c}\tilde{A}_{a}\right).
\end{align}
 We define its field strength as
\begin{align}
 Y^{\mathrm{p}}_{ab}\equiv2\nabla_{[a}X_{b]}^{\mathrm{p}}=\intv\tilde{Y}^{\mathrm{p}}_{ab}\ipq,\label{eq:Y_p}
\end{align}
where $\tilde{Y}^{\mathrm{p}}$ is
\begin{align}
    \tilde{Y}^{\mathrm{p}}_{ab}\equiv2\ii\left(k_{[a}+l_{[a}\right)\tilde{X}_{b]}^{\mathrm{p}}.
\end{align}

The homogeneous solution $X^{\mathrm{h}}$ is governed by the wave equation $\nabla^{b}\nabla_{b}X^{\mathrm{h}}_{a}=0$. Accordingly, we consider a solution form of
\begin{align}
    X^{\mathrm{h}}_{a}\left(t,\bm{x}\right)=\int d^{2}\mu\int_{\substack{-\infty\\\fho\neq0}}^{\infty}\frac{d\fho}{2\pi}\tilde{X}^{\mathrm{h}}_{a}\left(\fho,\mu\right)e^{\ii R\left(t,\bm{x};\fho,\mu\right)},\label{eq:X_h}
\end{align}
where $\mu$ is the spatial unit vector, $d^{2}\mu$ is the solid angle element of three-dimensional unit sphere, $\fho$ is the frequency, $\tilde{X}^{\mathrm{h}}$ is the amplitude of $\left(\fho,\mu\right)$, and $R\left(t,\bm{x};\fho,\mu\right)\equiv\fho\left(-t+\mu\cdot\bm{x}\right)$ is the phase. The field strength of the homogeneous solution $Y^{\mathrm{h}}_{ab}$ is defined by
\begin{align}
    Y^{\mathrm{h}}_{ab}\equiv2\nabla_{[a}X^{\mathrm{h}}_{b]}=\int d^{2}\mu\int_{\substack{-\infty\\\fho\neq0}}^{\infty}\frac{d\fho}{2\pi}\tilde{Y}^{\mathrm{h}}_{ab}\left(\fho,\mu\right)e^{\ii R},
\end{align}
where the amplitude $\tilde{Y}^{\mathrm{h}}_{ab}$ is determined by a boundary condition.

We impose a boundary condition on a plane that is perpendicular to $\lambda$ and located at the source of the EMW such that the field strength perturbation $Y_{ab}=Y^{\mathrm{p}}_{ab}+Y^{\mathrm{h}}_{ab}$ vanishes. It means that all geodesic observers located on the plane emit electromagnetic waves identical to \cref{eq:4_potential} in their own frame. Explicitly, the boundary condition is described as $Y_{ab}\left(t,\bm{y}\right)=0$ for all $\left(t,\bm{y}\right)$ such that $\lambda\cdot\bm{y}=-d$ where $d$ is distance between the source plane and the origin. Assuming that a receiver is in the positive $\lambda$ side from the plane, directions of waves $\mu$ incident to the receiver from the source plane is restricted to $\mu\cdot\lambda>0$. We only consider these waves. As shown by Appendix \ref{app:homogeneous_solution}, the solution that meets all conditions we imposed is given by
\begin{align}
    Y^{\mathrm{h}}_{ab}&=-\intv\tilde{Y}^{\mathrm{p}}_{ab}\ipqd,\label{eq:Y_h}
\end{align}
where
\begin{align}
    \Delta\left(\bm{x}\right)&\equiv\left\{\sqrt{\fe^{2}+2\fg\fe+\left(\fg\kappa^{a}\lambda_{a}\right)^{2}}\right.
    \nonumber\\&\qquad\qquad\left.-\fe-\fg\kappa^{a}\lambda_{a}\right\}\left(\lambda\cdot\bm{x}+d\right).\label{eq:Delta}
\end{align}

Collecting $Y^{\mathrm{p}}$ in \cref{eq:Y_p} and $Y^{\mathrm{h}}$ in \cref{eq:Y_h}, we obtain the field strength perturbation $Y$ as
\begin{align}
    Y_{ab}=\intv\tilde{Y}^{\mathrm{p}}_{ab}e^{\ii\left(P+Q\right)}\left(1-e^{\ii\Delta}\right).
\end{align}
Then, the perturbations of electric field and magnetic field are given by 
\begin{align}
    \delta E_{a}&=Y_{ab}n^{b},
    \\\delta B_{a}&=\frac{1}{2}\tensor{\varepsilon}{^{bc}_{a}}Y_{bc}-\tensor{\varepsilon}{^{bc}_{a}}\tensor{h}{^{d}_{b}}F_{dc}.
\end{align}
Finally, the perturbations of spatial scalars are obtained by
\begin{align}
    \delta I_{1}&=2g^{ab}\delta E_{a}E_{b}-h^{ab}E_{a}E_{b},\label{eq:I_1}
    \\\delta I_{2}&=2g^{ab}\delta B_{a}B_{b}-h^{ab}B_{a}B_{b},
    \\\delta I_{3}&=g^{ab}\delta E_{a}B_{b}+g^{ab}\delta B_{a}E_{b}-h^{ab}E_{a}B_{b}.
\end{align}

\section{Application to PTA}

For GW observation, we consider the square of the electric field $\eps{I}{_{1}}$. For example, a static conducting wire can only be powered by an electric field. Using three conducting wires oriented in different directions, we can measure three components of the electric field. Eventually, $\eps{I}{_{1}}$ is constructed by these components.

Let us consider an idealized pulsar rotating with exact frequency $\fz$. A plane EMW propagating to $\lambda$ from the pulsar has an electric field satisfying the periodic condition,
\begin{align}
    E_{a}\left(u+T\right)&=E_{a}\left(u\right),
\end{align}
where $u\left(t,\bm{x}\right)\equiv t-\lambda\cdot\bm{x}$ and $T\equiv2\pi/\fz$. Then, the electric field is decomposed into the Fourier series given by
\begin{align}
    E_{a}\left(u\right)&=\sum_{n=-\infty}^{\infty}\tilde{E}^{\left(n\right)}_{a}e^{-\ii n\fz u},\label{eq:E_pulse}
\end{align}
where $\tilde{E}^{\left(n\right)}$ are Fourier coefficients satisfying $\tilde{E}^{\left(-n\right)}=\tilde{E}^{\left(n\right)*}$. 
Meanwhile, in the case of an actual pulsar, the electric field will deviate from the above. We deal with this effect as an intrinsic noise. The derived noise in an observation can be reduced by a correlation method with a PTA, which will be introduced later.

A radio telescope measures an EMW with a frequency band from $100\mathrm{MHz}$ to $100\mathrm{GHz}$ (\cite{bolli_international_2019}). However, the frequency of pulsar $\fz$ is at most $\sim1\mathrm{KHz}$ (\cite{hessels_radio_2006}). Accordingly, we only consider the partial frequency range of EMW inside the radio telescope frequency band as
\begin{align}
    E_{a}\left(u\right)=\sum_{\left|n\right|=n_{1}}^{n_{2}}\tilde{E}^{\left(n\right)}_{a}e^{-\ii n\fz u},\label{eq:EMP_E}
\end{align}
where $n_{1}$ and $n_{2}$ are positive integers much larger than $1$, and the summation with vertical bars is defined by
\begin{align}
    \sum_{\left|n\right|=n_{1}}^{n_{2}}&\equiv\sum_{n=n_{1}}^{n_{2}}+\sum_{n=-n_{2}}^{-n_{1}}.
\end{align}

To introduce the geometrical optics, we need the condition of $\fe=n\fz\gg\fg$. In our situation, it is already the case that $n\gg1$. Therefore, GWs of frequencies comparable to $\fz$ are in the geometrical optics regime. Keeping the leading order $O\left(\fe/\fg\right)$ and the next order $O\left(1\right)$ in \cref{eq:Delta}, we obtain
\begin{align}
    \Delta\left(\bm{x}\right)\simeq\fg\left(1-\kappa^{a}\lambda_{a}\right)\left(\lambda\cdot\bm{x}+d\right).
\end{align}
Note that $\Delta$ does not depend on $\fe$ because the leading order vanishes. By substitution of \cref{eq:EMP_E} into \cref{eq:I_1}, keeping the the leading order and the next order as well, we obtain
\begin{align}
    \delta I_{1}&=-\int_{\kappa\neq\lambda}d^{2}\kappa\int_{\left|\fg\right|=\omega_{1}}^{\omega_{2}}\frac{d\fg}{2\pi}\sum_{\left|n\right|=n_{1}}^{n_{2}}\sum_{\left|m\right|=n_{1}}^{n_{2}}\tilde{E}^{\left(n\right)}_{a}\tilde{E}^{\left(m\right)}_{b}
    \nonumber\\&\times\left\{\frac{1}{\Theta}\left(1+n\frac{\fz}{\fg}\right)\tilde{h}_{cd}\lambda^{c}\lambda^{d}P_{\left(\lambda\right)}^{ab}\left(1-\id\right)+\tensor{\tilde{h}}{^{ab}}\id\right\}
    \nonumber\\&\times\ip e^{-\ii\left(n+m\right)\fz u},
\end{align}
where $\left(\omega_{1},\omega_{2}\right)$ is the GW frequency range, $\Theta\equiv1-\kappa^{a}\lambda_{a}$, and $P^{ab}_{\left(\lambda\right)}\equiv g^{ab}+n^{a}n^{b}-\lambda^{a}\lambda^{b}$ is the projection orthogonal to $n$ and $\lambda$. The integral with vertical bars is defined by
\begin{align}
    \int_{\left|\fg\right|=\omega_{1}}^{\omega_{2}}\frac{d\fg}{2\pi}&\equiv\int_{\omega_{1}}^{\omega_{2}}\frac{d\fg}{2\pi}+\int_{-\omega_{2}}^{-\omega_{1}}\frac{d\fg}{2\pi}.
\end{align}

\section{Frequency Window I}

We set a frequency window of range $\left(0,\fz/2\right)$ for $\eps{I}{_{1}}=\epsilon^{2}I_{1}+\epsilon^{3}\delta I_{1}+O\left(\epsilon^{4}\right)$. We also assume for simplicity that the GW frequency range is below to $\fz/2$. Then, we obtain $I_{1}$ and $\delta I_{1}$ as
\begin{align}
    I_{1}&=2P^{ab}\sum_{n=n_{1}}^{n_{2}}\tilde{E}^{\left(n\right)}_{a}\tilde{E}^{\left(n\right)*}_{b},\label{eq:I_1_w1}
    \\\delta I_{1}&=-2\int_{\kappa\neq\lambda}d^{2}\kappa\int_{\left|\fg\right|=\omega_{1}}^{\omega_{2}}\frac{d\fg}{2\pi}\sum_{n=n_{1}}^{n_{2}}\tilde{E}^{\left(n\right)}_{a}\tilde{E}^{\left(n\right)*}_{b}
    \nonumber\\&\times\left\{\frac{1}{\Theta}\tilde{h}_{cd}\lambda^{c}\lambda^{d}P^{ab}_{\left(\lambda\right)}\left(1-\id\right)+\tensor{\tilde{h}}{^{ab}}\id\right\}\ip.\label{eq:delta_I_w1}
\end{align}
Note that $\delta I_{1}$ is gauge-invariant because $I_{1}$ is constant scalar. This statement is proved by the lemma in \cite{stewart1974}.

To give an understandable picture, we simplify the EMW as a monochromatic plane wave of $\left(\fe,\lambda\right)$ having the form of
\begin{align}
    E_{a}=2\Re\left\{\left(\mathcal{E}_{\left(p\right)}p_{a}+\ii\mathcal{E}_{\left(s\right)}s_{a}\right)e^{\ii\left(Q+\delta_{\mathrm{e}}\right)}\right\},\label{eq:E_monochromatic_in_form}
\end{align}
where $\delta_{\mathrm{e}}$ is determined to make $\mathcal{E}_{\left(p\right)}$ and $\mathcal{E}_{\left(s\right)}$ real, and to make $\left\{n,p,s,\lambda\right\}$ an orthonormal basis. The ellipticity of the EMW polarization ellipse is defined by
\begin{align}
    \tan\chi\equiv\frac{\mathcal{E}_{\left(s\right)}}{\mathcal{E}_{\left(p\right)}}.
\end{align}
Likewise, an amplitude of GW $\tilde{h}_{ab}\left(\fg,\kappa\right)$ has the form of
\begin{align}
    \tilde{h}_{ab}=\left(\mathcal{H}_{+}e^{+}_{ab}+\ii\mathcal{H}_{\times}e^{\times}_{ab}\right)e^{\ii\delta_{\mathrm{g}}},\label{eq:h_in_form}
\end{align}
where $\delta_{\mathrm{g}}$ is determined to make $\mathcal{H}_{+}$ and $\mathcal{H}_{\times}$ real, and to make $\left\{e^{+},e^{\times}\right\}$ an orthonormal basis satisfying
\begin{align}
    0&=e^{A}_{ab}n^{b},\label{eq:e_n}
    \\0&=e^{A}_{ab}\kappa^{b},\label{eq:e_kappa}
    \\\delta^{AB}&=e^{A}_{ab}e^{B}_{cd}g^{ac}g^{bd},\label{eq:e_orthonormality}
\end{align}
for $A,B\in\left\{+,\times\right\}$. The set of all possible orthonormal bases orthogonal to $\kappa$ is parameterized by $\psi$ as
\begin{align}
    e^{+}_{ab}\left(\psi\right)&=\cos\left(2\psi\right)e^{+}_{ab}\left(0\right)+\sin\left(2\psi\right)e^{\times}_{ab}\left(0\right),
    \\e^{\times}_{ab}\left(\psi\right)&=-\sin\left(2\psi\right)e^{+}_{ab}\left(0\right)+\cos\left(2\psi\right)e^{\times}_{ab}\left(0\right),
\end{align}
where $\left\{e^{+}\left(0\right),e^{\times}\left(0\right)\right\}$ is a reference basis. We define $\theta$ and $\phi$ as the polar and azimuthal angle, respectively, of $\kappa$ with respect to the spatial frame $\left\{p,s,\lambda\right\}$ such that $\kappa^{a}=\sin\theta\cos\phi p^{a}+\sin\theta\sin\phi s^{a}+\cos\theta\lambda^{a}$.

\begin{figure}
    \centering
    \includegraphics[width=0.45\textwidth]{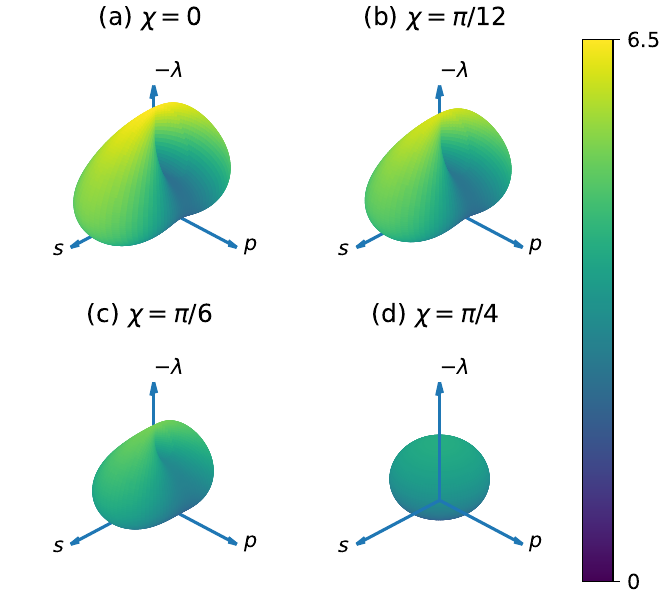}
    \caption{Antenna patterns $\left<\mathcal{F}\right>_{\psi,\Delta}$ of a GW with frequency window I in the spatial frame $\left\{\hat{x}=s,\hat{y}=p,\hat{z}=-\lambda\right\}$ where $\hat{z}$ is the direction of pulsar location. The incoming direction of a GW with polar and azimuthal angle $\left(\vartheta,\varphi\right)$ is related to $\left(\theta,\phi\right)$ by $\vartheta\equiv\theta$ and $\varphi\equiv-\pi/2-\phi$. For different EMW ellipticies $\chi$, (a) is the linear polarization directed to $p$, (b) and (c) are the elliptic polarizations, and (d) is the circular polarization. As seen in (a), a GW coming to $s$ orthogonally to polarization direction $p$ has more gain than a GW coming to $p$. For the circular polarization in (d), the gains for GWs coming to $p$ and $s$ are equal. The overall gain is largest in the linear polarization and decreases toward the circular polarization in accordance with \cref{eq:F_sky}.}
    \label{fig:antenna_pattern}
\end{figure}

Substituting \cref{eq:E_monochromatic_in_form,eq:h_in_form} into \cref{eq:I_1_w1,eq:delta_I_w1}, the amplitude of observation $\tilde{H}\left(\fg,\kappa\right)$ and the complex detector tensor $\tilde{D}^{ab}\left(\fg,\kappa\right)$  are defined as
\begin{align}
    \frac{\delta{I}_{1}}{I_{1}}&=\int_{\kappa\neq\lambda}d^{2}\kappa\int_{\left|\fg\right|=\omega_{1}}^{\omega_{2}}\frac{d\fg}{2\pi}\tilde{H}\ip,\label{eq:observtaion_H}
    \\\tilde{H}&=\tilde{h}_{ab}\tilde{D}^{ab},
    \\\tilde{D}^{ab}&=-\left(\cos^{2}\chi p^{a}p^{b}+\sin^{2}\chi s^{a}s^{b}\right)\id
    \nonumber\\&\qquad-\frac{1}{\Theta}\lambda^{a}\lambda^{b}\left(1-\id\right).
\end{align}
We consider the complex pattern functions $\tilde{F}^{A}$ for $A\in\left\{+,\times\right\}$ from
\begin{align}
    \tilde{H}=\mathcal{H}_{A}\tilde{F}^{A},
\end{align}
where we used the Einstein summation convention for index $A$. The strength of observation $\left|\tilde{H}\right|^{2}$ is given by
\begin{align}
    \left|\tilde{H}\right|^{2}=\mathcal{H}_{A}\mathcal{H}_{B}\mathcal{F}^{AB},
\end{align}
where $\mathcal{F}^{AB}\equiv\Re\left(\tilde{F}^{A}\tilde{F}^{*B}\right)$. To see the angular dependency of the strength, we consider the angle average of $|\tilde{H}|^{2}$ over $\Delta$ and $\psi$ fixing $\mathcal{H}_{A}$ because they can be any values from different pulsar locations and GW sources. As a result, we found that
\begin{align}
    \left<\mathcal{F}^{AB}\right>_{\psi,\Delta}=\frac{1}{2}\delta^{AB}\left<\mathcal{F}\right>_{\psi,\Delta},
\end{align}
where $\left<\cdot\right>_{\psi,\Delta}$ is the successive angle average over $\psi$ and $\Delta$, and $\mathcal{F}\equiv\delta^{AB}\mathcal{F}_{AB}$ is the trace of $\mathcal{F}^{AB}$. Angular dependencies of $\left<\mathcal{F}\right>_{\psi,\Delta}$ for different ellipticities $\chi$ are given by Appendix \ref{app:angular_dependency} and  depicted in \cref{fig:antenna_pattern}. The sky average of the strength is given by
\begin{align}
    \left<\left|\tilde{H}\right|^{2}\right>_{\mathrm{sky}}&=\frac{1}{2}\delta^{AB}\mathcal{H}_{A}\mathcal{H}_{B}\left<\mathcal{F}\right>_{\mathrm{sky}},
    \\\left<\mathcal{F}\right>_{\mathrm{sky}}&=\frac{11}{6}\left(1+\frac{3}{55}\cos\left(4\chi\right)-\frac{10}{11}\cos\Delta\right),\label{eq:F_sky}
\end{align}
where $\left<\cdot\right>_{\mathrm{sky}}$ is the successive angle average over $\theta$, $\phi$, and $\psi$.

We consider the observation of SGWs by the correlation between two observations from two different pulsars. Assuming that an SGW is stationary, Gaussian, isotropic, and evenly polarized, the correlation between two amplitudes in the SGW is given by
\begin{align}
    \left<\tilde{h}^{*}_{ab}\left(\fg,\kappa\right)\tilde{h}_{cd}\left(\fg',\kappa'\right)\right>&=2\pi\delta\left(\fg-\fg'\right)S_{h}\left(\fg\right)
    \nonumber\\&\times\frac{1}{4\pi}\delta^{2}\left(\kappa,\kappa'\right)P^{\left(\kappa\right)}_{abcd},
\end{align}
where $\left<\cdot\right>$ is the ensemble average, $S_{h}\left(\fg\right)$ is the power spectral density of the SGW, $P^{\left(\kappa\right)}_{abcd}\equiv\delta_{AB}e^{A}_{ab}\left(\psi\right)e^{B}_{cd}\left(\psi\right)$, and $\delta^{2}\left(\kappa,\kappa'\right)\equiv\delta\left(\cos\theta-\cos\theta'\right)\delta\left(\phi-\phi'\right)$. Although we used the orthonormal basis $\left\{e^{+}\left(\psi\right),e^{\times}\left(\psi\right)\right\}$ that depends on $\psi$, $P^{\left(\kappa\right)}$ is independent of the choice of $\psi$. 

In order to consider a practical observation, we introduce a noise term $N$ explicitly in the observation $\delta I_{1}/I_{1}$ at the origin as
\begin{align}
    \frac{\delta I_{1}}{I_{1}}&=H\left(t,\bm{x}=0\right)+N\left(t\right),
\end{align}
where $H\left(t,\bm{x}\right)$ is right-hand side of \cref{eq:observtaion_H}. The noise $N$ comes from various sources, e.g., the intrinsic noise of the pulsar, the fluctuation of pulse by the interstellar medium, and the detector noise. We assume that the noise $N$ is stationary Gaussian and has zero expectation that is achieved by a value shift of the observation. Also, we suppose that noises from two observations are uncorrelated. Then, the correlation between two observations is simplified as
\begin{align}
    \left<\frac{\delta I_{1}}{I_{1}}\frac{\delta I'_{1}}{I'_{1}}\right>&=\int_{\left|\fg\right|=\omega_{1}}^{\omega_{2}}\frac{d\fg}{2\pi}S_{h}\int_{\kappa\neq\lambda,\lambda'}\frac{d^{2}\kappa}{4\pi}\tilde{D}^{*ab}\tilde{D}'^{cd}P^{\left(\kappa\right)}_{abcd},
\end{align}
where quantities with prime are defined from a different pulsar. Notice that two pulsars have different frequencies $\fz$ and $\fz'$. Therefore, we need restriction of the frequency window by $\left(0,\min\left(\fz,\fz'\right)/2\right)$. Accordingly, the observational frequency of GWs is also limited to the same range. In the case of PTA, the observational frequency range for GWs is determined by the lowest rotational frequency among pulsars.

Assuming, $\fg d\gg1$, $\fg d'\gg1$, and that distance between two pulsars is much larger than wavelength of the SGW, the correlation at the origin approximately reduces to
\begin{align}
    \left<\frac{\delta I_{1}}{I_{1}}\frac{\delta I'_{1}}{I'_{1}}\right>&\simeq C\left(\lambda,\lambda'\right)\int_{\left|\fg\right|=\omega_{1}}^{\omega_{2}}\frac{d\fg}{2\pi}S_{h},
\end{align}
where $C\left(\lambda,\lambda'\right)$ is given by
\begin{align}
    C\left(\lambda,\lambda'\right)&\equiv\int_{\kappa\neq\lambda,\lambda'}\frac{d^{2}\kappa}{4\pi}\frac{1}{\Theta\Theta'}P^{\left(\kappa\right)}_{abcd}\lambda^{a}\lambda^{b}\lambda'^{c}\lambda'^{d}.
\end{align}
Having $\cos\alpha\equiv\lambda^{a}\lambda'_{a}$, $C$ becomes
\begin{align}
    C\left(\alpha\right)=1+\frac{1}{3}\cos\alpha+4\left(1-\cos\alpha\right)\ln\left(\sin\frac{\alpha}{2}\right),
\end{align}
which is identical to the \cite{hellings_upper_1983} curve up to factor of 4.

\section{Frequency Window II} 

We set a frequency window of range $\left(\fz/2,3\fz/2\right)$ for $\eps{I}{_{1}}$. We also assume that the GW frequency range is below to $\fz/2$. Then, we obtain $I_{1}$ and $\delta I_{1}$ as
\begin{align}
    I_{1}&=4\Re\left(P^{ab}\sum_{n=n_{1}}^{n_{2}-1}\tilde{E}^{\left(n\right)*}_{a}\tilde{E}^{\left(n+1\right)}_{b}e^{-\ii\fz u}\right),
    \\\delta I_{1}&=-2\Re\left[\int_{\kappa\neq\lambda}d^{2}\kappa\int_{\left|\fg\right|=\omega_{1}}^{\omega_{2}}\frac{d\fg}{2\pi}\sum_{n=n_{1}}^{n_{2}-1}\tilde{E}^{\left(n\right)*}_{a}\tilde{E}^{\left(n+1\right)}_{b}\right.
    \nonumber\\&\times\left.\left\{\frac{1}{\Theta}\left(1+\frac{\fz}{\fg}\right)\tilde{h}_{cd}\lambda^{c}\lambda^{d}P_{\left(\lambda\right)}^{ab}\left(1-\id\right)+2\tensor{\tilde{h}}{^{ab}}\id\right\}\right.
    \nonumber\\&\times\left.\ip e^{-\ii\fz u}\right].
\end{align}
Note that $\delta I_{1}$ is not gauge-invariant because $I_{1}$ is oscillating with frequency $\fz$.

To find a gauge-invariant quantity, we consider peak time $t_{k}$ of $I_{1}$ for integer $k$ satisfying
\begin{align}
    \dot{I}_{1}\left(t_{k}\right)=0,
\end{align}
where dot is the derivative $n^{a}\nabla_{a}$ and $t_{k}=t_{0}+kT$ because of the periodicity. Referring to Appendix \ref{app:peak_time_perturbation}, the perturbation of the peak time is given by
\begin{align}
    \delta t_{k}=-\frac{\delta\left(\dot{I}_{1}\right)\left(t_{k}\right)}{\ddot{I}_{1}\left(t_{k}\right)},\label{eq:delta_t_k}
\end{align}
where $\delta\left(\dot{I}_{1}\right)$ is easily obtained from the identity $\delta\left(\dot{I}_{1}\right)=n^{a}\nabla_{a}\delta I_{1}$ because $\delta n^{a}=0$. Then, quantity $\delta T_{k}=\delta t_{k+1}-\delta t_{k}$ is gauge-invariant because its background value is constant $T$. Because of its complexities, we think that $\delta T_{k}$ does not have many advantages compared to \cref{eq:delta_I_w1} in frequency window I. However, in limit $\fg\ll\fz$, it is simplified drastically and becomes
\begin{align}
    \frac{\delta T_{k}}{T}&=-\int_{\kappa\neq\lambda}d^{2}\kappa\int_{\left|\fg\right|=\omega_{1}}^{\omega_{2}}\frac{d\fg}{2\pi}\frac{\tilde{h}_{ab}\lambda^{a}\lambda^{b}}{2\Theta}
    \nonumber\\&\qquad\times\left(1-\id\right)e^{-\ii\fg\left(t_{k}-\kappa\cdot\bm{x}\right)},
\end{align}
which is used in the current GW observation by PTA.

\section{Conclusions}

We have obtained the perturbation of a plane EMW with arbitrary GWs. In the analysis, we directly solved the vacuum Maxwell equation imposing the boundary condition on the plane perpendicular to the propagating direction of the EMW located at a pulsar. This result was applied to the received EMW in a radio telescope, which partially composes the periodic EMP emitted by the pulsar. In the process, we assumed the geometrical optics such that the frequency of the received EMW is much larger than those of the EMP and the GWs. Because we did not impose the smallness of the GW frequency compared to the EMP frequency, the observational frequency range of PTA is extended by our results.

For the observation of GW, we introduced two frequency windows. In frequency window I, we proposed a gauge-invariant quantity that was the square of the electric field, and showed its antenna patterns as being different from the ellipticity of EMW. The correlation between observations from two pulsars has an angular dependency identical to the one given by Hellings and Downs. In frequency window II, we proposed a gauge-invariant quantity that was the time interval between adjacent peaks. Due to its complexity, there is no advantage over the gauge-invariant quantity in frequency window I. However, when the GW frequency is much smaller than the frequency of the pulsar, it is simplified considerably and becomes the formula used in the current GW observation by PTA.

\begin{acknowledgments}
C.P. was supported by the National Research Foundation of Korea (NRF) funded by the Korea government (NRF-2018R1D1A1B07041004, NRF-2020R1I1A2054376, NRF-2019R1A2C2006787) and by the National Institute for Mathematical Sciences (NIMS) funded by the Ministry of Science and ICT (B21710000).
\end{acknowledgments}

\appendix

\section{Homogeneous Solution of the perturbed Maxwell Equation}
\label{app:homogeneous_solution}

From the form of homogeneous solution $X^{\mathrm{h}}$ in \cref{eq:X_h}, the field strength $Y^{\mathrm{h}}$ is given by
\begin{align}
    Y^{\mathrm{h}}_{ab}&=2\nabla_{[a}X^{\mathrm{h}}_{b]}
    \nonumber\\&=\int_{\mu\cdot\lambda\geq0}d^{2}\mu\int_{\substack{-\infty\\\fho\neq0}}^{\infty}\frac{d\fho}{2\pi}\tilde{Y}^{\mathrm{h}}_{ab}\left(\fho,\mu\right)e^{\ii R\left(t,\bm{x};\fho,\mu\right)},\label{eq:Y_h_explicit}
\end{align}
where $\tilde{Y}^{\mathrm{h}}_{ab}=2\ii\mu_{[a}\tilde{X}^{\mathrm{h}}_{b]}$. Then, the boundary condition becomes
\begin{align}
    0&=Y_{ab}^{\mathrm{p}}\left(t,\bm{y}\right)+Y_{ab}^{\mathrm{h}}\left(t,\bm{y}\right)
    \nonumber\\&=\intv\tilde{Y}^{\mathrm{p}}_{ab}\ipq+\int_{\mu\cdot\lambda\geq0}d^{2}\mu\int_{\substack{-\infty\\\fho\neq0}}^{\infty}\frac{d\fho}{2\pi}\tilde{Y}^{\mathrm{h}}_{ab}\ir,
\end{align}
for all $\left(t,\bm{y}\right)$ such that $\lambda\cdot\bm{y}=-d$. We introduce a spatial Cartesian coordinate $\left\{x,y,z\right\}$ such that unit vector $\hat{z}$ is identical to $\lambda$. Therefore, the boundary is described by
\begin{align}
    \bm{y}=x\hat{x}+y\hat{y}-d\hat{z}.
\end{align}
By the Fourier transformation for variables $\left(t,x,y\right)$, we obtain $\tilde{Y}^{\mathrm{h}}$ as
\begin{align}
    \tilde{Y}_{ab}^{\mathrm{h}}\left(\fho,\mu\right)&=-\left(\frac{\fho}{2\pi}\right)^{2}\lambda\cdot\mu e^{\ii\fho\lambda\cdot\mu d}\int_{-\infty}^{\infty}dte^{\ii\fho t}
    \nonumber\\&\qquad\times\int_{-\infty}^{\infty}dxe^{-\ii\fho\hat{x}\cdot\mu x}\int_{-\infty}^{\infty}dye^{-\ii\fho\hat{y}\cdot\mu y}
    \nonumber\\&\qquad\times\intv\tilde{Y}^{\mathrm{p}}_{ab}e^{\ii\fg\left(-t+\kappa\cdot\bm{y}\right)}e^{\ii\fe\left(-t-d\right)}.
\end{align}
Using the Fourier transformation of the Dirac delta function, $\tilde{Y}^{\mathrm{h}}$ becomes
\begin{align}
    \tilde{Y}_{ab}^{\mathrm{h}}&=-2\pi\fho^{2}\lambda\cdot\mu\intv\tilde{Y}^{\mathrm{p}}_{ab}e^{-\ii\fg\kappa\cdot\lambda d}e^{-\ii\fe d}e^{\ii\fho\lambda\cdot\mu d}
    \nonumber\\&\qquad\times\delta\left(\fho-\fg-\fe\right)
    \nonumber\\&\qquad\times\delta\left(\hat{x}\cdot\left(\fg\kappa-\fho\mu\right)\right)\delta\left(\hat{y}\cdot\left(\fg\kappa-\fho\mu\right)\right),
\end{align}
where $\delta$ is Dirac delta function. By substitution of the above into \cref{eq:Y_h_explicit}, we obtain \cref{eq:Y_h}.

\section{Antenna Pattern of the Observation with Frequency Window I}
\label{app:angular_dependency}

\begin{align}
    \left<\mathcal{F}\right>_{\psi,\Delta}&=\frac{1}{2}\left\{\cos^{4}\chi-2\left(2\cos^{4}\chi-\cos^{2}\chi\right)\sin^{2}\phi\right.
    \nonumber\\&\qquad\left.+\left(4\cos^{4}\chi-4\cos^{2}\chi+1\right)\sin^{4}\phi\right\}\sin^{4}\theta
    \nonumber\\&-\left[\left\{\left(2\cos^{2}\chi-1\right)\sin^{2}\phi-\cos^{2}\chi\right\}\cos\theta\right.
    \nonumber\\&\qquad\left.+2\cos^{4}\chi-2\cos^{2}\chi+1\right]\sin^{2}\theta
    \nonumber\\&+\left\{2\left(2\cos\chi^{2}-1\right)\sin^{2}\phi-2\cos^{2}\chi+3\right\}\cos\theta
    \nonumber\\&+2\left(2\cos^{2}\chi-1\right)\sin^{2}\phi
    \nonumber\\&+2\cos^{4}\chi-4\cos^{2}\chi+\frac{7}{2}
\end{align}

\section{Perturbation of Peak Time in the Observation with Frequency Window II}
\label{app:peak_time_perturbation}

In the perturbed spacetime, $\eps{t}{_{k}}$ is determined from
\begin{align}
    \eps{\dot{I}}{_{1}}\left(\eps{t}{_{k}}\right)&=0.
\end{align}
Then,
\begin{align}
    0&=\eps{\dot{I}}{_{1}}\left(\eps{t}{_{k}}\right)-\dot{I}_{1}\left(t_{k}\right)
    \nonumber\\&=\eps{\dot{I}}{_{1}}\left(\eps{t}{_{k}}\right)-\eps{\dot{I}}{_{1}}\left(t_{k}\right)+\eps{\dot{I}}{_{1}}\left(t_{k}\right)-\dot{I}_{1}\left(t_{k}\right)
    \nonumber\\&=\eps{\ddot{I}}{_{1}}\left(t_{k}\right)\left(\eps{t}{_{k}}-t_{k}\right)+\epsilon\delta\left(\dot{I}_{1}\right)\left(t_{k}\right)+O\left(\epsilon^{2}\right)
    \nonumber\\&=\epsilon\left(\ddot{I}_{1}\left(t_{k}\right)\delta t_{k}+\delta\left(\dot{I}_{1}\right)\left(t_{k}\right)\right)+O\left(\epsilon^{2}\right).
\end{align}
From above, we obtain \cref{eq:delta_t_k}.

\bibliography{references}{}
\bibliographystyle{aasjournal}

\listofchanges

\end{document}